\documentclass[12pt,draftcls,onecolumn]{IEEEtran}

\IEEEoverridecommandlockouts                              
\usepackage{tipa}
\usepackage{pifont}
\usepackage{cases}
\usepackage{txfonts}
\usepackage{graphicx}
\usepackage{array}
\usepackage{mathrsfs}
\usepackage{amsfonts}
\usepackage{setspace}
\usepackage{subfigure}

\usepackage{setspace}
\newtheorem{Theorem}{Theorem}
\newtheorem{Lemma}{Lemma}
\newtheorem{Proposition}{Proposition}
\newtheorem{Remark}{Remark}
\newtheorem{Definition}{Definition}

\title{\LARGE \bf
Computation for Supremal Simulation-Based Controllable and Strong
Observable Subautomata }

\author{Yajuan Sun, Hai Lin, Fuchun Liu
\thanks{Yajuan Sun, Hai Lin are with Electrical and Computer Engineering Dept., National
University Of Singapore, 117576, Singapore (email: \{sunyajuan, elelh\}@nus.edu.sg)}%
\thanks{Fuchun Liu is with Faculty of Computer, Guangdong University of Technology, Guangzhou 510006, China (email: liufch8@gmail.com)}%
}

\begin{document}

\maketitle
\thispagestyle{empty}

\begin{abstract}
Bisimulation relation has been successfully applied to computer
science and control theory. In our previous work, simulation-based
controllability and simulation-based observability are proposed,
under which the existence of bisimilarity supervisor is
guaranteed. However, a given specification automaton may not
satisfy these conditions, and a natural question is how to compute
a maximum permissive sub-specification. This paper aims to answer
this question and investigate the computation of the supremal
simulation-based controllable and strong observable subautomata
with respect to given specifications by the lattice theory. In
order to achieve the supremal solution, three monotone operators,
namely simulation operator, controllable operator and strong
observable operator, are proposed upon the established complete
lattice. Then, inequalities based on these operators are
formulated, whose solution is the simulation-based controllable
and strong observable set. In particular, a sufficient condition
is presented to guarantee the existence of the supremal
simulation-based controllable and strong observable subautomata.
Furthermore, an algorithm is proposed to compute such subautomata.
\end{abstract}


\section{INTRODUCTION}
Bisimulation relation was introduced in \cite{milner} as a
behavioral equivalence relationship between two dynamical systems,
and since then it has been used widely in the study of discrete
event systems (DESs) \cite{f}, linear systems \cite{pa},
probabilistic systems \cite{pro}, and hybrid systems \cite{hyb}.
Bisimulation provides a stronger equivalence than the extensively
studied language equivalence \cite{cc}. It is known that the
language generated by two bisimilar systems are equivalent, but
the systems possessing the same language might not be bisimilar.
Moreover, two bisimilar systems have equivalent reachability
properties, or more generally, preserve properties specified in
terms of temporal logic such as CTL* \cite{tl}. Therefore, the
bisimilarity control that aims to achieve a bisimulation
equivalence between controlled system and specification has
attracted lots of attentions these years.

Komenda and Schuppen characterized the language controllability
and observability in terms of partial bisimulation by using
coalgebra for supervisory control of DESs under partial
observation \cite{k1}. Tabuada investigated the controller
synthesis problem of affine systems for bisimulation equivalence
\cite{TabA} and extended it to various systems including
discrete-event systems, nonlinear control systems, behavioral
systems, and hybrid systems by means of category theory
\cite{TabC}. In Zhou's work \cite{cbis3} and our previous work
\cite{liu}, the problem addressed is to design a supervisor to
execute the control action to achieve the bisimulation relation
between supervised system and specification, where plant and
specification are generally described as nondeterministic
automata. In Zhou's work \cite{cbis3}, a small model theorem is
established to show that the supervisor exists if and only if it
exists over the power set of Cartesian product of system and
specification state spaces.

In our previous work \cite{liu}, a different framework is proposed
to characterize the existence of the supervisor. The supervisor
exists if and only if the specification is simulation-based
controllable under full observation. As for the partial
observation case, the specification should be both
simulation-based controllable and simulation-based observable to
ensure the existence of the supervisor. However, in most
situations, a given specification does not satisfy those
conditions. Then, a natural question is how to compute a maximum
permissive sub-specification. Here, we would like to calculate the
supremal simulation-based controllable and strong observable
subautomata. Please note that the existing work for the
calculation of supremal controllable/normal sublanguages are all
based on the language controllability/normality \cite{ke},
\cite{kf}. To our best knowledge, there is no work considering the
computation of the supremal subautomata under simulation-based
controllability and simulation-based observability, where the
specifications are given as automata instead of languages.

This paper aims to answer this question and investigate the
computation of the supremal simulation-based controllable and
strong observable subautomata with respect to given specifications
by the lattice theory. Some preliminary results on the computation
of the supremal simulation-based controllable subautomata under
full observations were presented in \cite{sun}. In this paper, we
will calculate the supremal simulation-based controllable and
strong observable subautomata for the partial observation case. In
order to achieve the supremal solution, three monotone operators,
namely simulation operator, controllable operator and strong
observable operator, are proposed upon the established complete
lattice. Then, inequalities based on these three operators are
formulated, whose solution is the simulation-based controllable
and strong observable set. In particular, a sufficient condition
is presented to guarantee the existence of the supremal
simulation-based controllable and strong observable subautomata.
Furthermore, an algorithm is proposed to compute such subautomata.

This note is organized as follows. Section 2 gives the
preliminary. Section 3 reviews the works that have been done under
full observation. Section 4 studies the computation of the
supremal simulation-based controllable and strong observable
subautomata under partial observation. An illustrative example is
provided in Section 5. The note concludes with section 6.

\section{Preliminary}
\subsection{Discrete Event System}
A DES is modeled as an automaton $G =(X,\Sigma,x_{0},\alpha,
X_{m})$, where $X$ is the set of states, $\Sigma$ is a finite set
of events, $\alpha: X \times \Sigma \rightarrow 2^X$ is the
transition function, $x_0$ is the initial state, $X_m \subseteq X$
is the set of marked states. $\Gamma: X \rightarrow 2^{\Sigma}$ is
the active function and $\Gamma(x)$ is the active event set at
state $x$. Let $\Sigma^{*}$ be the set of all finite strings over
$\Sigma$, including the empty string $\epsilon$. Then the
transition function $\alpha$ can be extended to $\alpha: X \times
\Sigma^{*} \rightarrow 2^{X}$ in the nature way \cite{cc}. The
language generated by $G$ is defined as $L(G)=\{s \in \Sigma^{*}
\mid \alpha(x_0, s)$ is defined$\}$. The event set can be
partition into $\Sigma$ = $\Sigma_{uc}\dot{ \cup} \Sigma_{c}$,
where $\Sigma_{uc}$ is the set of uncontrollable events and
$\Sigma_{c}$ is the controllable event set. It can be also
partitioned into $\Sigma$ = $\Sigma_{uo} \dot{\cup} \Sigma_{o}$,
where $\Sigma_{uo}$ is the set of unobservable events and
$\Sigma_{o}$ is the set of observable events. Given an event
string $s \in \Sigma^*$, $|s|$ is the length of the string and
$s(i)$ is the $i^{th}$ event of this string, where $1 \leq i \leq
|s|$. When a string of events occurs, the sequence of observable
events is filtered by a projection $P$: $\Sigma^{*} \rightarrow
\Sigma_{o}^{*}$, which is defined inductively as follows:
$P(\epsilon)=\epsilon$, for $\sigma \in \Sigma$ and $s \in
\Sigma^{*}$, $P(s\sigma)=P(s)\sigma$ if $\sigma \in \Sigma_o$,
otherwise, $P(s\sigma)=P(s)$. The accessible operator $Ac$ is used
to remove the states which are not accessible from the initial
state, and it is defined as below.



\begin{Definition}
Given an automaton $G =(X,\Sigma,x_{0},\alpha,X_{m})$, the
accessible operator on G is defined as:
\[
Ac(G) = (X_{ac},\Sigma,x_{0},\alpha_{ac},X_{acm}),
\]
where $X_{ac} = \{ x \in X \mid  x \in \alpha_{ac}(x_0, s)$, where
$s \in \Sigma^{*}$ \}, $X_{acm} = X_m \cap X_{ac}$, $\alpha_{ac}$:
$X_{ac}\times \Sigma \rightarrow X_{ac}$ is a transition function,
and for any $x \in X_{ac}$ and $e \in \Sigma$, $\alpha_{ac}(x, e)
= \{y \in X_{ac} \mid y \in \alpha(x, e)\}$.
\end{Definition}

Further, the concept of subautomaton is introduced and a
subautomaton operator is proposed to construct a subautomaton from
a given state set.

\begin{Definition}
Given an automaton $G =(X,\Sigma,x_{0},\alpha,X_{m})$, the
subautomaton of $G$ is defined as $G_1 = (X_1, \Sigma_1, x_{0},
\alpha, X_{m1})$, where $X_1\subseteq X$, $X_{m1}\subseteq X_{m}$,
and $\alpha_{1}$ = $\alpha \mid X_{1}\times \Sigma \rightarrow
X_{1}$.
\end{Definition}
The notation $\alpha \mid X_{1}\times \Sigma \rightarrow X_{1}$
means that we are restricting $\alpha$ to the smaller domain of
the states $X_1$. The subautomaton of $G$ picks its states and
marked states from the corresponding sets in $G$.

\begin{Definition}
Given an automaton $R =(Q,\Sigma,q_{0},\delta,Q_{m})$, the
subautomata operator is defined as:
\[
Rc(Z) = Ac(Q_{rc},\Sigma,q_{0},\delta_{rc},Q_{rcm}),
\]
where Z $\subseteq Q \times X$, $Q_{rc}$ = \{$q \in Q \mid (q, x)
\in Z$\}, $Q_{rcm}$ = $Q_m \cap Q_{rc}$, and $\delta_{rc} = \delta
\mid Q_{rc} \times \Sigma \rightarrow Q_{rc}$.
\end{Definition}
By this subautomata operator, we can construct a subautomata of
the original automata $R$ from a set $Z$, whose elements are the
state pairs of $R$ and $G$. In addition, the state set $Q_{rc}$ of
this subautomata is a subset of the corresponding state set $Q$ of
$R$ and the transition function of this subautomata restricts
$\delta$ to a smaller domain of the states $Q_{rc}$.

Then, simulation relation is used to describe the equivalence
between automata as follows.
\begin{Definition}
Let $G_{1} =(X_{1},\Sigma,x_{01},\alpha_{1},X_{m1})$ and $G_{2}
=(X_{2},\Sigma,x_{02},\alpha_{2},X_{m2})$ be two automata. $G_{1}$
is said to be simulated by $G_{2}$, denoted by $G_{1} \prec_{\phi}
G_{2}$, if there is a binary relation $\phi$ $\subseteq$ $X_{1}
\times X_{2}$ such that $(x_{01},x_{02}) \in \phi$ and for each
$(x_{1}, x_{2}) \in \phi$,

(1) $x_{1}^{'} \in \alpha_{1}(x_{1},\sigma)$, where $\sigma \in
\Sigma \Rightarrow \exists x_{2}^{'} \in \alpha_{2}(x_{2},\sigma)$
such that $(x_{1}^{'},x_{2}^{'}) \in \phi$.

(2) $x_{1} \in X_{m1} $, then $x_{2} \in X_{m2} $.

\end{Definition}
If $G_{1} \prec_{\phi} G_{2}$, $G_{2} \prec_{\phi} G_{1}$, and
$\phi$ is symmetric, $\phi$ is a bisimulation relation between
$G_{1}$ and $G_{2}$, denoted by $G_{1} \simeq_{\phi} G_{2}$. We
sometimes omit the subscript $\phi$ from $\prec_{\phi}$ or
$\simeq_{\phi}$ when it is clear from the context. Moreover, the
main result of \cite{liu} is as below.

\begin{Theorem}\label{tliu}
Given a plant $G=(X,\Sigma,\alpha,x_{0},X_{m})$, a specification
$R=(Q,\Sigma,\delta,q_{0},Q_{m})$ and a projection $P$, assume
that $L(R)$ is language controllable and language observable.
Then, there exists a simulation relation $\Phi\subseteq Q\times X$
and a $P$-supervisor $S_{P}$ such that $S_{P}^{\Phi}/G\simeq R$
and $S_{P}^{\Phi}/G$ is $\Sigma_{uc}$-consistent if and only if
$R$ is simulation-based controllable and simulation-based
observable.
\end{Theorem}

The simulation-based controllability and simulation-based
observability are defined as below.


\begin{Definition}
Given a plant $G =(X,\Sigma,x_{0},\alpha,X_{m})$ and a
specification $R =(Q,\Sigma,q_{0},\delta,Q_{m})$, $R$ is
simulation-based controllable with respect to $G$ and
$\Sigma_{uc}$ if it satisfies:

(1) (Simulation Condition) There is a simulation relation $\phi$
such that $R \prec_{\phi} G$.

(2) (Controllable Condition) ($\forall$ $s\in L(R)$)($\forall$ $q$
$\in$ $\delta(q_{0},s)$)($\forall$ $\sigma \in
\Sigma_{uc}$)[$s\sigma \in L(G) \Rightarrow \delta(q,\sigma) \neq
\emptyset]$.
\end{Definition}

The set $Q_1 \times X_1 \subseteq Q \times X$ is said to be a
simulation-based controllable set if $Q_1 \times X_1$ is a
simulation relation from $R$ to $G$ and $Rc(Q_1 \times X_1)$
satisfies the controllable condition.

\begin{Definition}
Given a plant $G =(X,\Sigma,x_{0},\alpha,X_{m})$ and a
specification $R =(Q,\Sigma,q_{0},\delta,Q_{m})$. $R$ is said to
be simulation-based observable with respect to $G$, $\Sigma_{c}$
and $P$, if it satisfies:

(1) (Simulation Condition) There is a simulation relation $\phi$
such that $R \prec_{\phi} G$.

(2) (Observable Condition) $\forall s$, $s' \in L(R)$ with $P(s) =
P(s')$ $(\forall q \in \delta(q_0, s))$ $(\forall \delta \in
\Sigma_{c})$ $[s'\sigma \in L(R)$ and $s\sigma \in L(G) \Rightarrow
\delta(q, \sigma) \neq \emptyset$].
\end{Definition}

Simulation-based controllability and simulation-based
observability implies language controllability and language
observability, but the reverse does not hold.

\subsection{Lattice Theory}



\begin{Definition}
Consider a set $X$ and a relation $R \subseteq X \times X$ over $X$.
$R$ is reflexive if for each $x \in X$, $(x, x) \in R$; it is
antisymmetric if $(x, y) \in R$ and $(y, x) \in R$ implies $x=y$; it
is transitive if $(x, y) \in R$ and $(y, z) \in R$ implies that $(x,
z) \in R$. The partial order relation, denoted by $\leq$, over $X$
is a reflexive, antisymmetric and transitive relation. The pair $(X,
\leq)$ is a poset.
\end{Definition}

\begin{Definition}
Consider a set $X$ and $Y \subseteq X$. $x \in X$ is said to be the
supremal of $Y$, denoted by $supY$ or $\sqcup Y$, if it satisfies :
(1) $\forall y \in Y$: $y \leq x$, (2) $[\forall y \in Y : y \leq z]
\Rightarrow [x \leq z]$. $x$ is said to be the infimal of $Y$,
denoted by $infY$ and $\sqcap Y$, if it satisfies: (1) $\forall y
\in Y$: $x \leq y$ (2) $\forall z \in X$ : $[\forall y \in Y : z
\leq y] \Rightarrow [z \leq x]$. The poset $(X, \leq)$ is called a
lattice if $supY$, $infY \in X$ for any finite $Y$. If $supY, infY
\in X$ for arbitrary $Y \subseteq X$, then $(X, \leq)$ is called a
complete lattice.
\end{Definition}

A poset may be a lattice, but it may have a set $Y$ of infinite
size for which $infY$ or $supY$ may not exist. However, $infY$ and
$supY$ exist for any $Y \subseteq X$ on a complete lattice.
Moreover, monotone functions and disjunctive functions are defined
over a complete lattice $(X, \leq)$.

\begin{Definition}
A function $f : X \rightarrow X$ is said to be monotone if for any
$x, y \in X$ : $[x \leq y] \Rightarrow [f(x) \leq f(y)]$. $f$ is
said to be disjunctive if for any $Y \subseteq X : f(\sqcup_{y \in
Y}Y) = \sqcup_{y \in Y} f(y)$.
\end{Definition}


Furthermore, the following lemmas are introduced to obtain the
supremal solution of the system of inequalities \cite{kb}.

\begin{Lemma}\label{l1}

Consider the system of inequalities \{$f_{i}(x) \leq
g_{i}(x)$\}$_{i \leq n}$ over a compete lattice (X, $\leq$). Let Y
= \{$y \in X \mid \forall i \leq n: f_{i}(y) \leq g_{i}(y)$\} be
the set of all solutions of the system of inequalities and $Y_{1}$
= \{$y \in X \mid h_{1}(y) = y$\} be the set of all fixed points
of $h_{1}$, where $h_{1} = \sqcap_{i \leq
n}f_{i}^{\bot}(g_{i}(y))$ and $f_{i}^{\bot}(g_{i}(y))$ is the
supremal solution of $f_{i}(x) \leq g_{i}(x)$. If $f_{i}$ is
disjunctive and $g_{i}$ is monotone, then $supY \in Y$, $supY_{1}
\in Y_{1}$, and $supY = supY_{1}$.
\end{Lemma}

\begin{Lemma}\label{l2}
Consider the inequalities \{$f_{i}(x) \leq g_{i}(x)\}_{i \leq n}$
and $Y = \{ y \in X \mid \forall i \leq n: f_{i}(y) \leq
g_{i}(y)$\}. If $f_{i}$ is disjunctive and $g_{i}$ is monotone,
$supY$ can be obtained by iterative computation: $y_{0} = sup X$,
$\forall k \geq 0, y_{k+1} = h_{1}(y_{k})$ until $y_{m+1} = y_m =
supY$.


\end{Lemma}

In this note, we focus on the computation of the supremal
simulation-based controllable and strong observable subautomaton
for the specification, which is not simulation-based controllable
and observable.

\section{Full Observation}
In this section, we establish a complete lattice over which the
constructed simulation operator, controllable operator and their
properties are reviewed \cite{sun}.


\begin{Definition} Given a plant $G =(X,\Sigma,x_{0},\alpha,X_{m})$ and
a specification $R =(Q,\Sigma,q_{0},\delta,Q_{m})$, the poset is
defined as ($2^{Q \times X}, \subseteq$).
\end{Definition}

It can be seen that this power set lattice ($2^{Q \times X},
\subseteq$) is built upon the state pairs from $R$ and $G$ and it
is a complete lattice \cite{kb}. Thus, supremal and infimal
defined with respect to a compete lattice are unique.

\begin{Remark}
An alternative poset can be a prelattice $(\mathcal{S}, \prec)$,
where $\mathcal{S}$:= \{$S' ~|~ (S' \prec R) \wedge
(S'$~is~simulation-based controllable and strong observable) \} is
a set of automata and $\prec$ is a simulation relation. However,
the supremal solution with respect to the prelattice
$(\mathcal{S}, \prec)$ is not unique because this simulation
relation over $\mathcal{S}$ is a preorder, which is transitive,
reflexive but not anti-symmetric.



\end{Remark}

Next, we introduce several operators defined over ($2^{Q \times
X}, \subseteq$).


\begin{Definition}
The simulation operator $F_s: 2^{Q \times X} \rightarrow 2^{Q \times X}$ defined by
$(q, x) \in F_s(Z)$, for $ Z \subseteq Q \times X$, if the following
conditions are satisfied:

1. $(q, x) \in Z$.

2. $ q^{'} \in \delta(q, \sigma)$ $\Rightarrow$ $[\exists x^{'}
\in \alpha(x, \sigma)]$ $[(q^{'}, x^{'}) \in Z]$.

3. $q \in Q_{m} $ $\Rightarrow$ $x \in X_{m} $.
\end{Definition}

The simulation operator evolves from a similar operator in
\cite{pov} and it has following properties. Their proofs can be
found in \cite{sun}.

\begin{Proposition}\label{ss}
Given a plant $G =(X,\Sigma,x_{0},\alpha,X_{m})$ and a specification
$R =(Q,\Sigma,q_{0},\delta,Q_{m})$, $\phi$ is a simulation relation
from $R$ to $G$ if and only if $\phi \subseteq F_s(\phi)$ and $(q_0,
x_0) \in \phi$.
\end{Proposition}

\begin{Proposition}\label{sm}
Given a plant $G =(X,\Sigma,x_{0},\alpha,X_{m})$, a specification
$R =(Q,\Sigma,q_{0},\delta,Q_{m})$ and the sets $Z, Z' \subseteq Q
\times X$, $F_s(Z) \subseteq F_s(Z ^{'})$ if $Z \subseteq Z ^{'}$.
\end{Proposition}

\begin{Theorem} \label{st}
Given a plant $G =(X,\Sigma,x_{0},\alpha,X_{m})$ and a
specification $R =(Q,\Sigma,q_{0},\delta,Q_{m})$, the supremal
simulation relation is the maximal fixed-point $Z$ of the operator
$F_s$ if $(q_0, x_0) \in Z$, where $Z \subseteq Q \times X$.
Moreover,
\[
F_s(Z) = \lim_ {i \rightarrow \infty}F_s^{i}(Q \times X),
\]
where $F_s^{0}(Q \times X) = Q \times X$ is an identity function,
and for each $i \geq 0$, $F_s^{i+1}(Q \times X) = F_s(F_s^{i}(Q
\times X))$.
\end{Theorem}

%
%


Before presenting the controllable operator, we introduce the
following concepts.
\begin{Definition}
Given an automaton $G_1 =(X_1,\Sigma_1,x_{0},\alpha_1,X_{1m},
\Gamma_1)$ and a state $x \in X_1$, the string set of $x$, denoted
by $S_{x}$, is defined as $S_{x}=\{s \in \Sigma^{*} ~|~ x \in
\alpha_1(x_0, s)\}$. The nondeterministic state set of $x$, denoted
by $X_{x}$, is defined as $X_{x}=\{ x \in X_1 ~|~ x \in
\alpha_1(x_0, s), s \in S_{x}\}$. Further, we define the
nondeterministic active event set of the state $x$, denoted by
$\Gamma_{n}(x)$, as $\Gamma_{n}(x) = \cup_{x_1 \in
X_{x}}\Gamma_1(x_1)$.
\end{Definition}

We can obtain all the strings that can reach $x$ from $x_0$ through
$S_x$ and all the states that are reachable from $x_{0}$ with the
strings in $S_{x}$ by $X_{x}$. Besides, $\Gamma_{n}(x)$ is a union
of the active event set of the states in $X_{x}$. Next, we propose
the following notion to guarantee the existence of the supremal
simulation-based controllable subautomata.
%
\begin{Definition}
Given a plant $G$ and a specification
$R=(Q,\Sigma,q_{0},\delta,Q_{m})$, $R$ is said to be calculable for
the supremal simulation-based controllable subautomaton with respect
to $G$ if it satisfies:
\[
(\forall q \in Q_{M})(\forall s \in S_q)(\forall \sigma \in
\Sigma_{uc})[s\sigma \in L(G) \Rightarrow \delta(q, \sigma) \neq
\emptyset]
\]
where $Q_{M}=\{q \in Q~|~|S_q| \geq 2\}$.
\end{Definition}

Before presenting the controller operator, the simulation-based
controllable product is established.
\begin{Definition}
Given a plant $G =(X,\Sigma,x_{0},\alpha,X_{m})$ and a specification
$R =(Q,\Sigma,q_{0},\delta,Q_{m})$, the simulation-based
controllable product of $R$ and $G$ is an automaton:
\[
R \times_{sc} G = Ac ( Q \times X \cup \{(q_{v}, x_{v})\}, \Sigma,
q_0 \times x_0, \gammaup_{sc}, Q_m \times X_m)
\]
where
\[
\ \gammaup_{sc}((q,x),\sigma) = \left\{ {\begin{array}{*{20}c}
   (q_{v}, x_{v}) & {\sigma \in (\Sigma_{uc} \cap (\Gamma_n(x) - \Gamma (q)) };  \\
   (\delta(q, \sigma), \alpha (x, \sigma)) & {\sigma \in \Gamma(x) \cap \Gamma(q)};  \\
   undefined & {\textrm{otherwise}}.  \\
\end{array}} \right.
\]
\end{Definition}
According to the definition of simulation-based controllable
product, a transition that leads to the new states through event
$\sigma$ is allowed if the active event sets of this state pair
$(q, x)$ share the event $\sigma$. Besides, there will be a
transition to $(q_{v}, x_{v})$ if the state $q$,  which is
reachable from initial state $q_0$ of $R$ along $s$, does not
include the uncontrollable event $\sigma$, where $\sigma$ is
defined at a certain state of $G$ reachable from its initial state
$x_0$ through $s$. Moreover, the state pairs that are not
reachable from $(q_0, x_0)$ are removed by the accessible
operator. Next, the controllable operator is built upon complete
lattice $(2^{Q \times X}, \subseteq)$.
\begin{Definition}
Given a plant $G =(X,\Sigma,x_{0},\alpha,X_{m})$, a specification $R
=(Q,\Sigma,q_{0},\delta,Q_{m})$ and an automaton $Rc(Z) \times_{sc}
G =(X_{scz}, \Sigma, q_0 \times x_0, \gammaup_{scz}, X_{sczm})$ for
$Z \subseteq Q \times X$, the controllable operator $F_c: 2^{Q
\times X} \rightarrow 2^{Q \times X}$ defined by $(q, x) \in F_c(Z)$
if it satisfies:
\[
(q, x)\notin Q_d(Z) \times X, Q_d(Z) =
\cup_{\sigma\in\Sigma_{uc}}Q_{d\sigma}(Z),
\]
where for any $\sigma\in\Sigma_{uc}$, $Q_{d\sigma}(Z)$ =
\{$q_{d\sigma} \in Q \mid (\exists x \in X)$ s.t. $(q_v, x_v) \in
\gammaup_{scz}((q_{d\sigma}, x), \sigma))$ \}.
\end{Definition}

Moreover, this controllable operator satisfies following properties.



\begin{Proposition} \label{cs}
Given a plant $G =(X,\Sigma,x_{0},\alpha,X_{m})$, a specification $R
=(Q,\Sigma,q_{0},\delta,Q_{m})$  and a set $Z \subseteq Q \times X$,
$Rc(Z)$ satisfies the controllable condition if $Z \subseteq F_c(Z)$
and there is $x\in X$ such that $(q_0, x) \in Z$.
\end{Proposition}

\begin{proof}
Assume that $Rc(Z)$ violates the controllable condition when $Z
\subseteq F_c(Z)$ and there is $x \in X$ such that $(q_0, x) \in Z$,
where $Z \subseteq Q \times X$, then there exists $s \in L(R)$ and
$\sigma \in \Sigma_{uc}$ such that $s\sigma \in L(G)$ and $q \in
\delta(q_0, s)$ with $\delta(q, \sigma) = \emptyset$. As $s\sigma
\in L(G)$, there is $x^{'}\in \alpha (x_0, s)$ with $\alpha(x',
\sigma) \neq \emptyset$. Moreover, $(q, x')$ belongs to the state
set of $Rc(Z) \times_{sc} G$ because it is reachable from $(q_0,
x_0)$ by the string $s$. Furthermore, we have $\sigma \in
\Sigma_{uc} \cap (\Gamma_n(x')-\Gamma(q))$ as $\delta(q, \sigma) =
\emptyset$ and $\alpha(x', \sigma) \neq \emptyset$. Thus, $(q_v,
x_v) \in \gammaup_{sc}((q, x'), \sigma)$ in $Rc(Z) \times_{sc} G$ by
the definition of the simulation-based controllable product. We
obtain $q \in Q_d(Z)$, therefore, $(q, x') \in Q_d(Z) \times X$. On
the other hand, we have $(q, x') \in F_c(Z)$ as $Z \subseteq
F_c(Z)$. Then, we obtain $(q, x')\notin Q_d(Z) \times X$ by the
definition of the controllable operator. Thus, there is a
contradiction. Therefore, $Rc(Z)$ satisfies the controllable
condition.
\end{proof}

\begin{Proposition}\label{cm}
Given a plant $G =(X,\Sigma,x_{0},\alpha,X_{m})$, a specification $R
=(Q,\Sigma,q_{0},\delta,Q_{m})$ and a set $Z \subseteq Q \times X$,
$F_c(Z) \subseteq F_c(Z')$ if $Z \subseteq Z'$ and $R$ is calculable
for the supremal simulation-based controllable subautomaton with
respect to $G$.
\end{Proposition}

\begin{proof}
For any $(q, x) \in F_c{(Z})$, we have $(q, x) \in Z$ and $(q, x)
\notin Q_d(Z) \times X$. Then, $(q, x)\in Z'$ since $Z \subseteq
Z'$. Further, $(q, x) \notin Q_d(Z') \times X$ because of the
definition of $Rc(Z') \times_{sc} G$ and the calculability of $R$
for the supremal simulation-based controllable subautomaton with
respect to $G$. Thus, $(q, x) \in F_c(Z')$. Therefore, we have
$F_c(Z) \subseteq F_c(Z')$.
\end{proof}

%
%

\section{Partial Observation}
In this section, we establish a monotone strong observable operator
over complete lattice $(2^{Q \times X}, \subseteq)$. Combine it with
the simulation operator and the controllable operator, the
inequalities whose solution is the simulation-based controllable and
strong observable set are set up. Then, an algorithm is proposed for
the computation of simulation-based controllable and strong
observable subautomata.

\subsection{Strong Observable Operator}

%

\begin{Definition}
Given a plant $G =(X,\Sigma,x_{0},\alpha,X_{m})$ and a
specification $R =(Q,\Sigma,q_{0},\delta,Q_{m})$, the
simulation-based observable product of $R$ and $G$ is defined as:
\[
R \times_{so} G = Ac(Q \times X, \overline{\Sigma}, q_0 \times x_0,
\gammaup_{so}, Q_m \times X_m)
\]
where $ \overline{\Sigma} = \Sigma\cup\{\epsilon\}$ and for any
$(q, x) \in Q \times X$, $(\sigma_1, \sigma_2) \in
\overline{\Sigma} \times \overline{\Sigma}$,

The transition $\gammaup_{so}((q,x), (\sigma_1,\sigma_2))$
\[
= \left\{ {\begin{array}{*{20}c}
   ( q, \alpha(x, \sigma_2))  &{P(\sigma_2)=\epsilon=\sigma_1};  \\
   (\delta(q, \sigma_1), x)  & {P(\sigma_1)=\epsilon=\sigma_2};  \\
   (q, x)  & {\sigma_1=\sigma_2=\epsilon};  \\
   (\delta(q, \sigma_1), \alpha (x, \sigma_2))  & {~~~~~(P(\sigma_1)=P(\sigma_2)) \wedge \!(\sigma_1 \neq \epsilon) \wedge (\sigma_2 \neq \epsilon) };  \\
   undefined  &{otherwise}. \\
\end{array}} \right.
\]
\end{Definition}

In particular, the transition can be extended from domain $Q
\times X \times  \overline{\Sigma} \times  \overline{\Sigma}$ to
domain $Q \times X \times \Sigma^{*} \times \Sigma^{*}$ in the
following recursive manner: $\gammaup_{so}((q,x),
(s_1\sigma_1,s_2\sigma_2))=
\gammaup_{so}(\gammaup_{so}(\gammaup_{so}((q,x), (s_1,s_2)),
(\epsilon, \sigma_2)), (\sigma_1, \epsilon))\cup \gammaup_{so}$
$(\gammaup_{so}(\gammaup_{so}((q,x), (s_1,s_2)), (\sigma_1,
\epsilon)), (\epsilon, \sigma_2)) $ $\cup
\gammaup_{so}(\gammaup_{so}((q,x),$ $(s_1,s_2)), (\sigma_1,
\sigma_2))$ if $\sigma_1, \sigma_2 \in \Sigma_{uo}$, otherwise,
$\gammaup_{so}((q,x), (s_1\sigma_1,s_2\sigma_2))=
\gammaup_{so}(\gammaup_{so}((q,x), (s_1,s_2)), (\sigma_1,
\sigma_2))$.

The simulation-based observable product $R \times_{so} G$ satisfies
the following proposition.

\begin{Proposition} \label{all}
Given a plant $G =(X,\Sigma,x_{0},\alpha,X_{m})$, a specification
$R =(Q,\Sigma,q_{0},\delta,Q_{m})$ and their simulation-based
observable product $R \times_{so} G = (X_{so}, \overline{\Sigma},
q_0 \times x_0, \gammaup_{so}, X_{som})$, $(q, x) \in
\gammaup_{so}((q_0, x_0), (s, s'))$ iff there exists $s, s'$ with
$P(s)=P(s')$ such that $q \in \delta(q_0, s)$ and $x \in
\alpha(x_0, s')$.
\end{Proposition}

\begin{proof}
The induction method is adopted to prove this proposition.
(Necessity) 1. $|s|=0$, then $s=\epsilon$. (1) $|s'|=0$, that is,
$s'=\epsilon$. Let $(q,x) \in \gammaup_{so}((q_0, x_0), (\epsilon,
\epsilon))$. Obviously, we have $q \in \delta(q_0, \epsilon)$, $x
\in \alpha(x_0, \epsilon)$ and $P(\epsilon)=P(\epsilon)$. (2) Let
$|s'|=1$ with $s'=\sigma_1$. For any $ (q',x') \in
\gammaup_{so}((q_0, x_0), (\epsilon, \sigma_1))$, we have
$P(\sigma_1)=\epsilon$, $q' \in \delta(q_0, \epsilon)$ and $x' \in
\alpha(x_0, \sigma_1)$. (3) Assume that $|s'|=n_2$, the necessity
of this proposition holds. (4) $|s'|=n_2+1$. For any $(q_1, x_1)
\in \gammaup_{so}((q_0, x_0), (\epsilon, s''\sigma_2))$, where
$s'=s''\sigma_2$, there exists $(q_2, x_2) \in \gammaup_{so}((q_0,
x_0), (\epsilon, s''))$ with $q_2 \in \delta(q_0, \epsilon)$, $x_2
\in \alpha(x_0, s'')$ and $P(s'')=\epsilon$ s.t. $(q_1, x_1) \in
\gammaup_{so}((q_2, x_2), (\epsilon, \sigma_2))$ since the
necessity of this proposition holds when $|\epsilon|=0$ and
$|s''|=n_2$. Then, $P(s''\sigma_2)=\epsilon$, $q_1 \in \delta(q_0,
\epsilon)$ and $x_1 \in \alpha(x_0, s')$. 2. Let $|s|=1$ with $s =
\sigma_3$. (1) $|s'|=0$, then $s'=\epsilon$. Obviously, the
necessity holds. (2) Let $|s'|=1$ with $s'=\sigma_4$. Any
$(q_3,x_3) \in \gammaup_{so}((q_0, x_0), (\sigma_3, \sigma_4))$
satisfies the following cases. Case $1$: there exists $(q_4,x_4)
\in \gammaup_{so}((q_0, x_0), (\epsilon, \sigma_4))$ with $q_4 \in
\delta(q_0, \epsilon)$, $x_4 \in \alpha(x_0, \sigma_4)$ and
$P(\sigma_4)=\epsilon$ s.t. $(q_3,x_3) \in \gammaup_{so}((q_4,
x_4), (\sigma_3, \epsilon))$, then
$P(\sigma_4)=\epsilon=P(\sigma_3)$, $q_3 \in \delta(q_0,
\sigma_3)$ and $x_3 \in \alpha(x_0, \sigma_4)$. Or case $2$: there
exists $(q_5,x_5) \in \gammaup_{so}((q_0, x_0), (\sigma_3,
\epsilon))$ with $q_5 \in \delta(q_0, \sigma_3)$, $x_5 \in
\alpha(x_0, \epsilon)$ and $P(\sigma_3)=\epsilon$ s.t. $(q_3,x_3)
\in \gammaup_{so}((q_5, x_5), (\epsilon, \sigma_4))$, then
$P(\sigma_3)=\epsilon=P(\sigma_4)$, $q_3 \in \delta(q_0,
\sigma_3)$ and $x_3 \in \alpha(x_0, \sigma_4)$. Or case $3$: there
exists $(q_3,x_3) \in \gammaup_{so}((q_0, x_0), (\sigma_3,
\sigma_4))$ then $P(\sigma_4)=P(\sigma_3)$, $\sigma_3 \neq
\epsilon$, $\sigma_4 \neq \epsilon$, $q_3 \in \delta(q_0,
\sigma_3)$ and $x_3 \in \alpha(x_0, \sigma_4)$. (3) Assume that
$|s'|=n_2$, the necessity of this proposition holds when $|s|=1$.
(4) $|s'|=n_2+1$. For any $(q_6, x_6) \in \gammaup_{so}((q_0,
x_0), (\sigma_3, s')$, where $s'=s'(1)\cdots s'(i)s'(i+1)\cdots
s'(|s'|-1)s'(|s'|))$, we have following cases. Case 1: there
exists $(q_7,x_7) \in \gammaup_{so}((q_0, x_0), (\sigma_3,
s'(1)\cdots s'(i)s'(i+1)\cdots s'(|s'|-1)))$ with $q_7 \in
\delta(q_0, \sigma_3)$, $x_7 \in \alpha(x_0, s'(1)\cdots
s'(i)s'(i+1)\cdots s'(|s'|-1))$ and $P(s'(1)\cdots
s'(|s'|-1))=P(\sigma_3)$ s.t. $(q_6,x_6) \in \gammaup_{so}((q_7,
x_7), (\epsilon, s'(|s'|)))$ since $|\sigma_3|=1$ and
$|s'(1)\cdots s'(i)s'(i+1)\cdots s'(|s'|-1)|=n_2$. Then,
$P(\sigma_3)=P(s')$, $q_6 \in \delta(q_0, \sigma_3)$ and $x_6 \in
\alpha(x_0, s')$. Or case 2: there exists $(q_8,x_8) \in
\gammaup_{so}((q_0, x_0), (\epsilon, s'))$ with $q_8 \in
\delta(q_0, \epsilon)$, $x_8 \in \alpha(x_0, s')$ and
$P(s')=\epsilon$ s.t. $(q_6,x_6) \in \gammaup_{so}((q_8, x_8),
(\sigma_3, \epsilon))$ since it is similar to 1.(4) when
$|\epsilon|=0$ and $|s'|=n_2+1$. Then,
$P(\sigma_3)=\epsilon=P(s')$, $q_6 \in \delta(q_0, \sigma_3)$ and
$x_6 \in \alpha(x_0, s')$. Or case 3: there exists $(q_9,x_9) \in
\gammaup_{so}((q_0, x_0), (\epsilon, s'(1)\cdots
s'(i)s'(i+1)\cdots s'(|s'|-1))$ with $q_9 \in \delta(q_0,
\epsilon)$, $x_9 \in \alpha(x_0, s'(1)\cdots s'(i)s'(i+1)\cdots
s'(|s'|-1))$ and $P(s'(1)\cdots s'(|s'|-1))=\epsilon$ s.t.
$(q_6,x_6) \in \gammaup_{so}((q_9, x_9), (\sigma_3, s'(|s'|))$
since it satisfies the case 1.(3) when $|\epsilon|=0$ and
$|s'(1)\cdots s'(i)s'(i+1)\cdots s'(|s'|-1)|=n_2$. Then,
$P(\sigma_3)=P(s')$, $q_6 \in \delta(q_0, \sigma_3)$ and $x_6 \in
\alpha(x_0, s')$. 3. Assume that $|s|=n_1$, $|s'|=n_2$, the
necessity of this proposition holds. (4) Let $|s|=n_1+1$ and
$|s'|=n_2$. For any $(q_{10}, x_{10}) \in \gammaup_{so}((q_0,
x_0), (s, s')$, it satisfies the following cases. Case 1: there
exists $(q_{11},x_{11}) \in \gammaup_{so}((q_0, x_0), (s(1)\cdots
s(i)s(i+1)\cdots s(|s|-1), s'(1)\cdots s'(i)s'(i+1)\cdots
s'(|s'|-1)))$ with $q_{11} \in \delta(q_0, s(1)\cdots
s(i)s(i+1)\cdots s(|s|-1))$, $x_{11} \in \alpha(x_0, s'(1)\cdots
s'(i)s'(i+1)\cdots s'(|s'|-1))$ and $P(s'(1)\cdots
s'(|s'|-1))=P(s(1)\cdots s(|s|-1))$ s.t. $(q_{10},x_{10}) \in
\gammaup_{so}((q_{11}, x_{11}), (s(|s|), s'(|s'|)))$ since the
necessity of this proposition holds when $|s(1)\cdots
s(i)s(i+1)\cdots s(|s|-1)|=n_1$ and $|s'(1)\cdots
s'(i)s'(i+1)\cdots s'(|s'|-1)|=n_2-1$. Then, $P(s)=P(s')$, $q_{10}
\in \delta(q_0, s)$ and $x_{10} \in \alpha(x_0, s')$. Or case 2:
there exists $(q_{12},x_{12}) \in \gammaup_{so}((q_0, x_0),
(s(1)\cdots s(i)s(i+1)\cdots s(|s|-1), s'))$ with $q_{12} \in
\delta(q_0, s(1)\cdots s(i)s(i+1)\cdots s(|s|-1))$, $x_{12} \in
\alpha(x_0, s')$ and $P(s')=P(s(1)\cdots s(|s|-1))$ s.t.
$(q_{10},x_{10}) \in \gammaup_{so}((q_{12}, x_{12}), (s(|s|),
\epsilon))$ since it satisfies 3 when $|s(1)\cdots
s(i)s(i+1)\cdots s(|s|-1)|=n_1$ and $|s'(1)\cdots
s'(i)s'(i+1)\cdots s'(|s'|-1)|=n_2$. Then, $P(s)=P(s')$, $q_{10}
\in \delta(q_0, s)$ and $x_{10} \in \alpha(x_0, s')$. Or case 3:
there exists $(q_{13},x_{13}) \in \gammaup_{so}((q_0, x_0), (s,
s'(1)\cdots s'(i)s'(i+1)\cdots s'(|s'|-1)))$ s.t. $(q_{10},x_{10})
\in \gammaup_{so}((q_{13}, x_{13}), (\epsilon, s'(|s'|)))$.
Similarly, we obtain $P(s)=P(s')$, $q_{10} \in \delta(q_0, s)$ and
$x_{10} \in \alpha(x_0, s')$. (Sufficiency) 1. $|s|=0$, then
$s=\epsilon$. Let $q \in \delta(q_0, \epsilon)$. (1) $|s'|=0$ and
$s'=\epsilon$. For any $x \in \alpha(x_0, \epsilon)$, it is
obvious that $(q,x) \in \gammaup_{so}((q_0, x_0), (\epsilon,
\epsilon))$. (2) $|s'|=1$. Let $s'=\sigma_1$ with
$P(\sigma_1)=\epsilon$. For any $ x' \in \alpha(x_0, \sigma_1)$,
we have $(q, x') \in \gammaup_{so}((q_0, x_0), (\epsilon,
\sigma_1))$. (3) Assume that the sufficiency of this proposition
holds when $|s|=0$ and $|s'|=n_2$. (4) $|s'|=n_2+1$. For any $x''
\in \alpha(x_0, s')$ with
$\epsilon=P(\epsilon)=P(s')=P(s'(1)\cdots s'(i)\cdots
s'(|s'|-1)\sigma_2)=P(s'(1)\cdots s'(i)\cdots
s'(|s'|-1))\sigma_2$, we obtain $P(\sigma_2)=\epsilon$. Because
the sufficiency of this proposition holds when $|s|=0$ and
$|s'(1)\cdots s'(i)\cdots s'(|s'|-1)|=n_2$ from above assumption,
there exists $x_1 \in \alpha(x_0, s'(1)\cdots s'(i)\cdots
s'(|s'|-1))$ with $x'' \in \alpha(x_1, \sigma_2)$ s.t. $(q, x_1)
\in \gammaup_{so}((q_0, x_0), (\epsilon, s'(1)\cdots s'(i)\cdots
s'(|s'|-1)))$, then $(q, x'') \in \gammaup_{so}((q_0, x_0),
(\epsilon, s'(1)\cdots s'(i)\cdots s'(|s'|-1))\sigma_2))$. 2.
$|s|=1$. Let $s=\sigma_3$ and $q_1 \in \delta(q_0, \sigma_3)$. (1)
$|s'|=0$ and $s'=\epsilon$. For any $x''' \in \alpha(x_0,
\epsilon)$ with $P(\sigma_3)=\epsilon=P(\epsilon)$, we have
$(q_1,x''') \in \gammaup_{so}((q_0, x_0), (\sigma_3, \epsilon))$.
(2) $|s'|=1$. Let $s'=\sigma_4$ with $P(\sigma_3)=P(\sigma_4)$,
$\sigma_3 \neq \epsilon$ and $\sigma_4 \neq \epsilon$. Then, for
any $ x_1 \in \alpha(x_0, \sigma_4)$, we have $(q, x_1) \in
\gammaup_{so}((q_0, x_0), (\sigma_3, \sigma_4))$. (3) Assume that
the sufficiency of this proposition holds when $|s|=1$ and
$|s'|=n_2$. (4) $|s'|=n_2+1$. Let $x_2 \in \alpha(x_0,
s')=\alpha(x_0, s'(1)\cdots s'(i)\cdots s'(|s'|-1)\sigma_4)$ with
$P(\sigma_3)=P(s')=P(s'(1)\cdots s'(i)\cdots s'(|s'|-1)\sigma_4)$.
If $P(\sigma_3)=\epsilon$, then $P(s'(1)\cdots s'(i)\cdots
s'(|s'|-1))=P(\sigma_4)=\epsilon=P(\sigma_3)$. There exists $x_3
\in \alpha(x_0, s'(1)\cdots s'(i)\cdots s'(|s'|-1))$ with $x_2 \in
\alpha(x_3, \sigma_4)$ s.t. $(q_1, x_3) \in \gammaup_{so}((q_0,
x_0), (\sigma_3, s'(1)\cdots s'(i)\cdots s'(|s'|-1)))$ as the
sufficiency of this proposition holds when $|\sigma_3|=1$ and
$|s'(1)\cdots s'(i)\cdots s'(|s'|-1)|=n_2$. Moreover, $(q_1, x_2)
\in \gammaup_{so}((q_1, x_3), (\epsilon, \sigma_4))$. Then, $(q_1,
x_2) \in \gammaup_{so}((q_0, x_0), (\sigma_3, s'))$. If
$P(\sigma_3)=\sigma_3 \neq \epsilon$, we have
$P(\sigma_3)=P(s'(1)\cdots s'(i)\cdots s'(|s'|-1)\sigma_4)$. Then,
there are two cases. Case 1: If
$P(\sigma_4)=P(\sigma_3)=\sigma_3$, we obtain $P(s'(1)\cdots
s'(i)\cdots s'(|s'|-1))=\epsilon$. Obviously, the sufficiency of
the proposition holds. Case 2: If $P(\sigma_4)=\epsilon$, there
exists $i \in N^{+}$ with $1\leq i \leq |s'|-1$ s.t.
$P(s'(i))=\sigma_3$ and $P(s'(1)s'(2)\cdots s'(i-1))=\epsilon$.
Futher, $P(s'(i+1)\cdots s'(|s'|-1)\sigma_4)=\epsilon$ if $1\leq i
< |s'|-1$; $P(\sigma_4)=\epsilon$ if $i=|s'|-1$. Thus, there is
$x_4 \in \alpha(x_0, s'(1)\cdots s'(i-1))$ with $x_5 \in
\alpha(x_4, \sigma_3)$ s.t. $(q_0, x_4) \in \gammaup_{so}((q_0,
x_0), (\epsilon, s'(1)\cdots s'(i-1)))$ as $1\leq|s'(1)\cdots
s'(i-1)|<n_2$. Then, $(q_1, x_5) \in \gammaup_{so}((q_0, x_4),
(\sigma_3, s'(i)))$ because of $P(s'(i))=\sigma_3$. Therefore, we
have $(q_1, x_2) \in \gammaup_{so}((q_0, x_0), (\sigma_3, s'))$ by
the definition of the simulation-based observable product. 3.
Assume that $|s|=n_1$, $|s'|=n_2$, the sufficiency of this
proposition holds. (4) $|s|=n_1+1$ and $|s'|=n_2$. Let $q_2 \in
\delta(q_0, s)=\delta(q_0, s(1)\cdots s(i)\cdots
s(|s|-1)\sigma_5)$, $x_6 \in \alpha(x_0, s')=\alpha(x_0,
s'(1)\cdots s'(i)\cdots s'(|s'|-1)\sigma_6)$ and $P(s)=P(s')$. If
$P(\sigma_5)=\epsilon$, then $P(s(1)\cdots s(i)\cdots
s(|s|-1))=P(s')$ with $|s(1)\cdots s(i)\cdots s(|s|-1)|=n_1$ and
$s'=n_2$. Thus, there exists $q_3 \in \delta(q_0, s(1)\cdots
s(i)\cdots s(|s|-1))$ with $q_2 \in \delta(q_3, \sigma_5)$ s.t.
$(q_3, x_6) \in \gammaup_{so}((q_0, x_0), (s(1)\cdots s(i)\cdots
s(|s|-1), s'))$. Therefore, $(q_2, x_6) \in \gammaup_{so}((q_3,
x_6), (\sigma_5, \epsilon))$. Then, $(q_2, x_6) \in
\gammaup_{so}((q_0, x_0), (s, s'))$. If $P(\sigma_5)=\sigma_5 \neq
\epsilon$, we have $P(s(1)\cdots s(i)\cdots
s(|s|-1))\sigma_5)=P(s(1)\cdots s(i)\cdots
s(|s|-1))\sigma_5=P(s'(1)\cdots s'(i)\cdots s'(|s'|-1)\sigma_6)$.
Then, we have two cases. Case 1: If $P(\sigma_6)=P(\sigma_5)$, we
obtain $P(s'(1)\cdots s'(i)\cdots s'(|s'|-1))= P(s(1)\cdots
s(i)\cdots s(|s|-1))$. There is $x_7 \in \alpha(x_0, s'(1)\cdots
s'(i-1))$ with $x_6 \in \alpha(x_7, \sigma_6)$ and $q_3' \in
\delta(q_0, s(1)\cdots s(i)\cdots s(|s|-1))$ with $q_2 \in
\delta(q_3', \sigma_5)$ s.t. $(q_3, x_7) \in \gammaup_{so}((q_0,
x_0), (s(1)\cdots s(i)\cdots s(|s|-1), s'(1)\cdots s'(i-1)))$
because the sufficiency of the proposition holds when $|s(1)\cdots
s(|s|-1)|=n_1$ and $|s'(1)\cdots s'(|s'|-1)|=n_2-1$. Then, $(q_2,
x_6) \in \gammaup_{so}((q_3', x_7), (\sigma_5, \sigma_6))$.
Therefore, $(q_2, x_6) \in \gammaup_{so}((q_0, x_0), (s, s'))$ by
the definition of the simulation-based observable product. Case 2:
If $P(\sigma_6)=\epsilon$. There exists $i$ with $1\leq
i\leq|s'|-1$ s.t. $P(s'(i))=\sigma_5$ with $P(s(1)s(2)\cdots
s(|s|-1))=P(s'(1)s'(2)\cdots s'(i-1))$. Moreover, $P(s'(i+1)\cdots
s'(|s'|-1)\sigma_6)=\epsilon$ if $1 \leq i < |s'|-1$ and
$P(\sigma_6)=\epsilon$ if $i=|s'|-1$. Thus, there is $x_8 \in
\alpha(x_0, s'(1)\cdots s'(i-1))$ with $x_9 \in \alpha(x_8,
\sigma_5)$ s.t. $(q_3, x_8) \in \gammaup_{so}((q_0, x_0),
(s(1)s(2)\cdots s(|s|-1), s'(1)\cdots s'(i-1)))$ as
$|s(1)s(2)\cdots s(|s|-1)|=n_1$ and $1\leq|s'(1)\cdots
s'(i-1)|<n_2$ satisfying the assumption 3. Then, $(q_2, x_6) \in
\gammaup_{so}((q_0, x_0), (s, s'))$ because of the definition of
the simulation-based observable product.
\end{proof}

Based on the simulation-based observable product, the following
concepts are introduced.

\begin{Definition}
Given a simulation-based observable product $R \times_{so} G =
(X_{so}, \overline{\Sigma}, q_0 \times x_0, \gammaup_{so},
X_{som})$ and $s_1 \in L(R)$, the equivalent projection string set
of $s_1$ with respect to the plant $G$ is defined as
$S_{s_1}=\{s_2 \in \Sigma^{*} ~| ~\exists ~(q, x) \in X_{so}$ s.t.
$(q, x)\in \gammaup_{so}((q_0, x_0),(s_1, s_2))\}$.
\end{Definition}

It can be seen that all the strings of plant $G$ that have the same
projection as the string $s_1$ of specification are included in
$S_{s_1}$. In order to guarantee the existence of the supremal
simulation-based strong observable subautomata, we propose the
following concept.

\begin{Definition}
Given a plant $G$ and a specification
$R=(Q,\Sigma,q_{0},\delta,Q_{m})$, $R$ is said to be calculable
for the supremal simulation-based strong observable subautomaton
with respect to $G$ if it satisfies:
\[
(\forall q \in Q_{M})(\forall s \in S_q)(\forall s' \in
S_{s})(\forall \sigma \in \Sigma_{c})[s'\sigma \in L(G)
\Rightarrow \delta(q, \sigma) \neq \emptyset]
\]
where $Q_{M}=\{q \in Q~|~ |S_q| \geq 2\}$.
\end{Definition}

The specification $R$ is said to be calculable for
simulation-based controllable and strong observable subautomaton
with respect to $G$ if it is calculable for both supremal
simulation-based controllable subautomaton and supremal
simulation-based strong observable subautomaton.

Because the simulation-based observability is not closed under
state union, the supremal simulation-based observable subautomaton
does not exist. Here, we introduce the simulation-based strong
observability which implies simulation-based observability and it
is also closed under state union under certain conditions.
\begin{Definition}
Given a plant $G =(X,\Sigma,x_{0},\alpha,X_{m})$, a specification
$R =(Q,\Sigma,q_{0},\delta,Q_{m})$ and their simulation-based
observable product $R \times_{so} G = (X_{so}, \overline{\Sigma},
q_0 \times x_0, \gammaup_{so}, X_{som})$, $R$ is said to be
simulation-based strong observable with respect to $G$,
$\Sigma_{c}$ and $P$ if it satisfies:

(1) (Simulation Condition) There is a simulation relation $\phi$
such that $R \prec_{\phi} G$.

(2) (Strong Observable Condition) $[(s = \epsilon)(\forall s' \in
S_{\epsilon}) \Rightarrow s' \in L(R)]$ and $(\forall s_1 \in L(R)
\backslash \{\epsilon\})$$(\forall s_2 \in S_{s_1})$ $(\forall q
\in \delta(q_0, s_2))$$(\forall \sigma \in \Sigma_{c})$
[$s_1\sigma, s_2\sigma \in L(G) \Rightarrow \delta(q,\sigma) \neq
\emptyset]$.
\end{Definition}

The $Q_1 \times X_1 \subseteq Q \times X$ is said to be a
simulation-based strong observable set if $Q_1 \times X_1$ is a
simulation relation from $R$ to $G$ and $Rc(Q_1 \times X_1)$
satisfies the strong observable condition. Furthermore, the set
$Q_1 \times X_1$ is a simulation-based controllable and strong
observable set if it is a simulation-based controllable set and
also a simulation-based strong observable set.

The relationship between simulation-based strong observability and
simulation-based observability as below.
\begin{Proposition}
Given a plant $G =(X,\Sigma,x_{0},\alpha,X_{m})$ and a
specification $R =(Q,\Sigma,q_{0},\delta,Q_{m})$, $R$ is
simulation-based observable with respect to $G$, $\Sigma_{c}$ and
$P$ if $R$ is simulation-based strong observable with respect to
$G$, $\Sigma_{c}$ and $P$.
\end{Proposition}

\begin{proof}
Because $R$ is simulation-based strong observable with respect to
$G$, $\Sigma_{c}$ and $P$, we have that $R$ is simulated by $G$.
Assume that $R$ satisfies the strong observable condition but not
the observable condition, then there exists $s, s' \in L(R)$ with
$P(s)=P(s')$ s.t. there is $q \in \delta(q_0, s)$ with $\delta(q,
\sigma)=\emptyset$ if $s\sigma \in L(G)$ and $s'\sigma \in L(R)$,
where $\sigma \in \Sigma_{c}$. Let $s = \epsilon$, we have the
following cases. (1) $s'=\epsilon$. Since $\epsilon\sigma \in
L(R)$, $\delta(q, \sigma) \neq \emptyset$. (2) $s' \neq \epsilon$
with $P(s')=\epsilon$. We have $s' \in L(R)$ and $\epsilon \in
S_{s'}$. Moreover, $\epsilon\sigma \in L(G)$ and $s'\sigma \in
L(G)$ because $R \prec G$ implies $L(R) \subseteq L(G)$. Thus,
$\delta(q, \sigma) \neq \emptyset$ according to the strong
observable condition. Let $s \in L(R) \backslash \{\epsilon\}$, we
have $s \in S_{s}$. In addition, $s\sigma \in L(G)$. Thus,
$\delta(q, \sigma) \neq \emptyset$ because $R$ satisfies the
strong observable condition. Therefore, all the cases contradict
the assumption. As a result, $R$ satisfies the observable
condition. Hence, $R$ is simulation-based observable with respect
to $G$, $\Sigma_c$ and $P$.
\end{proof}

Based on the simulation-based strong observability, we propose the
following notion.

\begin{Definition}
Let $G =(X,\Sigma,x_{0},\alpha,X_{m})$ be a plant, $R
=(Q,\Sigma,q_{0},\delta,Q_{m})$ be a specification, $R \times_{so}
G = (X_{so}, \overline{\Sigma}, q_0 \times x_0, \gammaup_{so},
X_{som})$ be their simulation-based observable product and $Rc(Z)
=(Q_{rcz},\Sigma,q_{0},\delta_{rcz},Q_{rczm})$ be a subautomaton
for $Z \subseteq Q \times X$ . For any $s_1 \in L(Rc(Z))$, $s_2
\in S_{s_1}$ and $\sigma \in \Sigma_{c}$, the state failure set of
$s_1$ for the strong observability, denoted by $Q_{ds_1}'(Z)$, is
defined as:
\[
Q_{ds_1}'(Z) = \left\{ {\begin{array}{*{20}c}
  \{q \in Q ~|~ q \in \delta_{rcz}(q_0, s_1) \wedge (s_2 \notin L(Rc(Z)))\} &{s_1=\epsilon};  \\
  \{q \in Q ~|~ q \in \delta_{rcz}(q_0, s_2) \wedge (\delta_{rcz}(q, \sigma) = \emptyset) \wedge (s_1\sigma, s_2\sigma \in L(G))\} & {s_1 \neq \epsilon}.  \\
\end{array}} \right.
\]
\end{Definition}

Then, we construct the strong observable operator based on the
complete lattice $(2^{Q \times X}, \subseteq)$.
\begin{Definition}
Given a plant $G =(X,\Sigma,x_{0},\alpha,X_{m})$, a specification
$R =(Q,\Sigma,q_{0},\delta,Q_{m})$ and a subautomaton $Rc(Z)
=(Q_{rcz},\Sigma,q_{0},\delta_{rcz},Q_{rczm})$ for $Z \subseteq Q
\times X$, the strong observable operator $F_{so}: 2^{Q \times X}
\rightarrow 2^{Q \times X}$ defined by $(q, x) \in F_{so}(Z)$ if
it satisfies:
\[
(q, x)\notin Q_d'(Z) \times X, Q_d'(Z) = \cup_{s_1 \in
L(Rc(Z))}Q_{ds_1}'(Z).
\]
\end{Definition}

The strong observable operator satisfies following propositions.

\begin{Proposition}\label{ns}
Given a plant $G =(X,\Sigma,x_{0},\alpha,X_{m})$, a specification
$R =(Q,\Sigma,q_{0},\delta,Q_{m})$ and a set $Z \subseteq Q \times
X$, $Rc(Z)$ satisfies the strong observable condition if $Z
\subseteq F_{so}(Z)$ and there exists $x \in X$ such that $(q_0,
x) \in Z$.
\end{Proposition}

\begin{proof}
Let $Rc(Z)=(Q_{rcz},\Sigma,q_{0},\delta_{rcz},Q_{rczm})$ be a
subautomaton for $Z$. Assume that $Rc(Z)$ violates the strong
observable condition when $Z \subseteq F_{so}(Z)$ and $(q_0, x)
\in Z$, where $x \in X$, then there exists $s_1 \in L(Rc(Z))
\backslash \{\epsilon\}, s_2 \in S_{s_1}, \sigma \in \Sigma_{c}$
with $s_1\sigma, s_2\sigma \in L(G)$ such that $(q, x) \in Z$ with
$q \in \delta_{rcz}(q_0, s_2)$ and $\delta_{rcz}(q, \sigma) =
\emptyset$. Thus, $q \in Q_{ds_1}'(Z)$. Then, $(q, x) \in
Q_{d}'(Z) \times X$. Since $Z \subseteq F_{so}(Z)$, we have $(q,
x) \in F_{so}(Z)$. By the definition of the strong observable
operator $F_{so}(Z)$, $(q, x) \notin Q_{d}'(Z) \times X$, which
introduces a contradiction. Then, $Rc(Z)$ satisfies the strong
observable condition.
\end{proof}

\begin{Proposition}\label{nm}
Given a plant $G =(X,\Sigma,x_{0},\alpha,X_{m})$, a specification $R
=(Q,\Sigma,q_{0},\delta,Q_{m})$ and the sets $Z, Z' \subseteq Q
\times X$, $F_{so}(Z) \subseteq F_{so}(Z')$ if $Z \subseteq Z'$ and
$R$ is calculable for the supremal simulation-based strong
observable subautomaton with respect to $G$.
\end{Proposition}

\begin{proof}
Let $Rc(Z)=(Q_{rcz},\Sigma,q_{0},\delta_{rcz},Q_{rczm})$ be a
subautomaton for $Z$ and
$Rc(Z')=(Q_{rcz'},\Sigma,q_{0},\delta_{rcz'}$,
\\
$Q_{rcz'm})$ be a subautomaton for $Z'$. For any $(q, x) \in Z$
and $(q, x) \in F_{so}(Z)$, we have $(q, x) \notin Q_d'(Z) \times
X$. If $q=q_0$, then $s' \in L(Rc(Z))$ for any $s' \in
S_{\epsilon}$. Thus, we obtain $s' \in L(Rc(Z'))$ because of
$L(Rc(Z))\subseteq L(Rc(Z'))$. Hence, $q \notin
Q_{d\epsilon}'(Z)$. Since $(q, x) \in F_{so}(Z)$, we also have $q
\in \delta_{rcz}(q_0, s_2)$ with $\delta_{rcz}(q, \sigma) \neq
\emptyset$ for any $s_1 \in L(Rc(Z)) \backslash \{\epsilon\}$ such
that $s_1\sigma \in L(G)$ and any $s_2 \in L(Rc(Z))$ such that
$s_2\sigma \in L(G)$ and $s_2 \in S_{s_1}$. Moreover, $(q ,x) \in
Z'$ as $Z\subseteq Z'$. Assume that there exists $s_3 \in
L(Rc(Z')) \backslash \{\epsilon \}$, $s_4 \in S_{s_3}$ and $\sigma
\in \Sigma_{c}$ with $s_3\sigma \in L(G)$ and $s_4\sigma \in L(G)$
such that $q \in \delta_{rcz'}(q_0, s_4)$ and $\delta_{rcz'}(q,
\sigma)=\emptyset$. If $s_3 \in L(Rc(Z))$, we have the following
cases: (1) $s_4 \in L(Rc(Z))$. Obviously, $\delta_{rcz'}(q,
\sigma) \neq \emptyset$. (2) $s_4 \notin L(Rc(Z))$, then
$\delta_{rcz'}(q, \sigma) \neq \emptyset$ since $R$ is calculable
for supremal simulation-based strong observable subautomaton with
respect to $G$. On the other side, there are two cases if $s_3
\notin L(Rc(Z))$. (1)$s_4 \in L(Rc(Z))$. Because $s_4 \in
P^{-1}[P(s_4)]$ and $s_4\sigma \in L(G)$, we obtain
$\delta_{rcz}(q, \sigma) \neq \emptyset$. Then, $\delta_{rcz'}(q,
\sigma) \neq \emptyset$. (2) $s_4 \notin L(Rc(Z))$. Because $R$ is
calculable for the supremal simulation-based strong observable
subautomaton with respect to $G$, we have $\delta_{rcz'}(q,
\sigma) \neq \emptyset$. Thus, we get $\delta_{rcz'}(q, \sigma)
\neq \emptyset$ from all above cases, which contradicts the
assumption that $\delta_{rcz'}(q, \sigma)=\emptyset$. Therefore,
$q \notin Q_{ds_3}'(Z')$. Hence, $(q, x) \notin Q_d'(Z') \times
X$. Similarly, we can prove that $(q, x) \in F_{so}(Z')$ when $q
\neq q_0$. As a result, $F_{so}(Z) \subseteq F_{so}(Z')$.
\end{proof}

From definition of $F_{so}(Z)$, we have $F_{so}(Z)\subseteq Z$.
Then, the supremal state set $Z$ satisfying $Z\subseteq F_{so}(Z)$
is a fixed point of $F_{so}$ from lattice theory. As $F_{so}$ is
monotone by Proposition \ref{nm}, the maximal fixed point of
$F_{so}$ can be obtained by iterating $F_{so}$, and it will be
discussed in next subsection.




\subsection{Supremal Simulation-based Strong Observable Subautomata}
A sufficient condition is proposed to guarantee the existence of
the supremal simulation-based strong observable set. Further, an
algorithm is presented to such subautomaton.


%

\begin{Proposition}\label{supsn}
Let $G =(X,\Sigma,x_{0},\alpha,X_{m})$ be a plant, $R
=(Q,\Sigma,q_{0},\delta,Q_{m})$ be a specification, $Y$ = \{ $Q_1
\times X_1 \subseteq {Q \times X} \mid (F(Q_1 \times X_1)
\subseteq F_s(Q_1 \times X_1)) \wedge (F(Q_1 \times X_1) \subseteq
F_{so}(Q_1 \times X_1))$ \} and $Y_2 = \{Q_1 \times X_1 \in 2^{Q
\times X} \mid h_2(Q_1 \times X_1)= Q_1 \times X_1 \}$ be a set of
fixed points of $h_2$. For any $Q_1 \times X_1 \in 2^{Q \times X}$
and identify function $F(Q_1 \times X_1)= Q_1 \times X_1$, the
function $h_2: 2^{Q \times X} \rightarrow 2^{Q \times X}$ is
defined as:
\begin{eqnarray}
h_2 \!(Q_1 \! \times \! X_1) &=& sup \! \{Q_2 \! \times \! X_2 \!
\in \! 2^{Q \times X}\! :\! F(Q_2 \times X_2) \! \subseteq \!
F_s(Q_1 \times X_1)\} \! \cap \! sup \! \{\! Q_3 \! \times \! X_3
\! \in \! 2^{Q \times X}\! : \! F(Q_3 \! \times \! X_3) \!
\subseteq \! F_{so}(Q_1 \! \times \! X_1)\} \nonumber
\end{eqnarray}
Then, any $Q_1 \times X_1 \in Y$ is a simulation-based strong
observable set and $supY = supY_2$ if $(q_0, x_0) \in Q_1 \times
X_1$ and $R$ is calculable for the supremal simulation-based
strong observable subautomaton with respect to $G$.
\end{Proposition}

\begin{proof}
As $(q_0, x_0) \in Q_1 \times X_1$ and $Q_1 \times X_1 \subseteq
F_s(Q_1 \times X_1)$, we obtain that $Q_1 \times X_1$ is a
simulation relation from $R$ to $G$ by Proposition \ref{ss}.
Moreover, $Rc(Q_1 \times X_1)$ satisfies the strong observable
condition by Proposition \ref{ns} because $(Q_1 \times X_1)
\subseteq F_{so}(Q_1 \times X_1)$. Hence, $Q_1 \times X_1$ is a
simulation-based strong observable set. From lattice theory,
($2^{Q \times X}$, $\subseteq$) is a compete lattice over which we
definite the simulation operator $F_s$ and the strong observable
operator $F_{so}$ which are monotone by Proposition \ref{sm} and
Proposition \ref{nm}. The identity function $F(Q_2 \times X_2)$ =
$Q_2 \times X_2$ and $F(Q_3 \times X_3)$ = $Q_3 \times X_3$ are
disjunctive. Hence, $supY = supY_2$ by Lemma 1.
\end{proof}

{\it Algorithm 1: } Given a plant $G =(X,\Sigma,x_{0},\alpha,X_{m})$
and a specification $R =(Q,\Sigma,q_{0},\delta,Q_{m})$, the
algorithm for computing the supremal simulation-based strong
observable subautomaton with respect to $G$, $\Sigma_c$, and $P$ is
as follows:

Step 1. Check whether $R$ is calculable for the supremal
simulation-based strong observable with respect to $G$. If not,
the supremal simulation-based strong observable subautomaton does
not exist, otherwise, go to step 2.

Step 2. Let $y_{0} = Q \times X$, $\forall l \geq 0$, $y_{l+1} =
h_{2}(y_{l})$ until $y_{l+1} = y_l$.

Step 3. If $(q_0, x_0) \notin y_l$, the supremal simulation-based
strong observable subautomaton does not exist, otherwise, if
$(q_0, x_0)$ $\in$ $y_l$, $Rc(y_l)$ is the supremal
simulation-based strong observable subautomaton with respect to
$G$, $\Sigma_c$, and $P$.


\begin{Remark}
Since $G$ and $R$ are nondeterministic, their number of
transitions are $O(|X|^2 \times |\Sigma|)$ and $O(|Q|^2 \times
|\Sigma|)$ respectively. So the complexity of the simulation-based
observable product is $O(|Q|^2 \times |X|^2 \times
|\Sigma+1|^{2})$. Then, the complexity of checking the
calculability of specification $R$ for the supremal
simulation-based strong observable subautomaton with respect to
$G$ is $O(|X|^{2} \times |\Sigma|+|Q|^2 \times |X|^2 \times
(|\Sigma|+1)^2)$. Further, the complexity of the simulation
operator is $O(|Q|^2 \times |X|^2 \times  |\Sigma|)$ and the most
iterative times is $|X| \times |Q|$, the complexity of Algorithm 1
is $O(|Q|^3 \times |X|^3 \times (|\Sigma|+1)^2)$.
\end{Remark}

\begin{Theorem}
Algorithm 1 is correct.
\end{Theorem}
\begin{proof}
We have $y_l = supY$ by Lemma 2 and Proposition \ref{supsn}.
Further, $y_l$ is a simulation-based strong observable set if
$(q_0, x_0) \in y_l$ and $R$ is calculable for the supremal
simulation-based strong observable subautomaton w.r.t $G$ by
Proposition \ref{supsn}. Therefore, $y_l$ is the supremal
simulation-based strong observable set. Base on it, we build the
subautomton $Rc(y_l)$. Therefore, $Rc(y_l)$ is the supremal
simulation-based strong observable subautomaton w.r.t. $G$,
$\Sigma_c$, and $P$.
\end{proof}

\subsection{Supremal Simulation-based Controllable and Strong Observable Subautomata}
Further, we propose a sufficient condition to guarantee the
existence of the supremal simulation-based controllable and strong
observable set and an algorithm to calculate such subautomaton.
\begin{Proposition}\label{supscn}
Let $G =(X,\Sigma,x_{0},\alpha,X_{m})$ be a plant, $R
=(Q,\Sigma,q_{0},\delta,Q_{m})$ be a specification, $Y$ = \{ $Q_1
\times X_1 \subseteq {Q \times X} \mid (F(Q_1 \times X_1)
\subseteq F_s(Q_1 \times X_1)) \wedge (F(Q_1 \times X_1) \subseteq
F_c(Q_1 \times X_1)) \wedge (F(Q_1 \times X_1) \subseteq
F_{so}(Q_1 \times X_1))$ \} and $Y_3 = \{Q_1 \times X_1 \in 2^{Q
\times X} \mid h_3(Q_1 \times X_1)= Q_1 \times X_1 \}$ is a set of
fixed points of $h_3$. For any $Q_1 \times X_1 \in 2^{Q \times X}$
and identify function $F(Q_1 \times X_1)= Q_1 \times X_1$, the
function $h_3: 2^{Q \times X} \rightarrow 2^{Q \times X}$ is
defined as:
\begin{eqnarray}
h_3(Q_1 \! \times \! X_1) &=& sup \{ Q_2 \times X_2 \in 2^{Q
\times X}:  F(Q_2 \times X_2) \subseteq  F_s(Q_1 \times X_1)\}
\cap  sup \{Q_3 \times X_3 \in 2^{Q \times X}: F(Q_3 \times X_3)
\subseteq \nonumber \\
& & F_c(Q_1 \times X_1)\} \cap  sup \{Q_4 \times X_4 \in 2^{Q
\times X}: F(Q_4 \times X_4) \subseteq F_{so}(Q_1 \times X_1)\}
\nonumber \
\end{eqnarray}
Then, any $Q_1 \times X_1 \in Y$ is a simulation-based
controllable and strong observable set and $supY = supY_3$ if
$(q_0, x_0) \in Q_1 \times X_1$ and $R$ is calculable for supremal
simulation-based controllable and strong observable subautomaton
with respect to $G$.
\end{Proposition}

{\it Algorithm 2:}  Given a plant $G =(X,\Sigma,x_{0},\alpha,X_{m})$
and a specification $R =(Q,\Sigma,q_{0},\delta,Q_{m})$, the
algorithm for computing the supremal simulation-based controllable
and strong observable subautomaton is as follows:

Step 1. Check whether $R$ is calculable for the supremal
simulation-based controllable and strong observable subautomaton
with respect to $G$. If not, the supremal simulation-based
controllable and strong observable subautomaton does not exist,
otherwise, go to step 2.

Step 2. Let $y_{0} = Q \times X$, $\forall n \geq 0$, $y_{n+1} =
h_{3}(y_{n})$ until $y_{n+1} = y_n$.

Step 3. If $(q_0, x_0) \notin y_n$, the supremal simulation-based
controllable and strong observable subautomaton does not exist,
otherwise, $Rc(y_n)$ is the supremal simulation-based controllable
and strong observable subautomaton if $(q_0, x_0) \in y_n$.


\begin{Remark}
The complexity of checking calculability of specification $R$ for
the supremal simulation-based controllable subautomaton is
$O(|X|^{2} \times |\Sigma| + |Q|^{2} \times |\Sigma|)$. Further,
the complexity of the Algorithm 1 and the simulation-based
controllable product are $O(|Q|^3\times |X|^3 \times
(|\Sigma|+1)^2)$ and $O(|Q|^2 \times |X|^2 \times |\Sigma|)$
respectively, the complexity of Algorithm 2 is $O(|Q|^3\times
|X|^3 \times (|\Sigma|+1)^2)$.
\end{Remark}

\begin{Theorem}
Algorithm 2 is correct.
\end{Theorem}

The proofs for Propositions \ref{supscn} and Theorem 4 are similar
to Proposition \ref{supsn} and Theorem 3.

\begin{Remark}
Since simulation-based strong observability implies
simulation-based observability, the supremal simulation-based
controllable and strong observable subautomaton is
simulation-based controllable and observable. Further, its
language is controllable and observable because simulation-based
controllability and observability implies language controllability
and observability \cite{liu}.
\end{Remark}

Further, this supremal controllable and strong observable
subautomaton satisfies the following property.


\begin{Proposition}\label{supa}
Given a specification $R=(Q, \Sigma, q_0, \delta, Q_m)$ and a
plant $G=(X, \Sigma, x_0, \alpha, X_m)$ such that $R\prec_{\phi}
G$, the subautomaton $R''=(Q'', \Sigma, q_0, \delta, Q_m'')$
obtained by Algorithm 2 is a supremal element of automata set
$\mathcal{S}$:= \{$S'~|$ ($S'$ $\prec R$) $\wedge$ ($S'$ ~is
~simulation-based controllable and strong observable) \} based on
the prelattice $(\mathcal{S}, \prec)$.
\end{Proposition}

\begin{proof}
Let $R'=(Q', \Sigma, q_0', \delta', Q_m')$ be an automaton
satisfies that $R'\prec_{\phi_{1}} R$ and $R'$ is simulation-based
controllable. We need to prove that there exists a simulation
relation $\phi_2$ between $R'$ and $R''$ such that $R'
\prec_{\phi_{2}} R''$ when $R\prec_{\phi} G$. Because
$R'\prec_{\phi_{1}} R$, there is $q_1 \in \delta(q_0, s_1)$ such
that $(q_1', q_1) \in \phi_{1}$ for any $q_1' \in \delta'(q_0',
s_1)$. Assume that $q_1 \in Q - Q''$, there are two cases
according to Algorithm 2: (1)($s_1\sigma\in L(G))\wedge(\sigma \in
\Sigma_{uc}$)$\wedge(\sigma \notin \Gamma(q_1))$. Then $\sigma
\notin \Gamma'(q_1')$ because $(q_1', q_1) \in \phi_{1}$ with
$\Gamma'(q_1')\subseteq \Gamma(q_1)$. Thus, $R'$ is not
simulation-based controllable w.r.t. $G$ and $\Sigma_{c}$, which
introduces a contradiction. (2) For any $s'$ such that $q_1 \in
\delta(q_0, s')$, we have any $q_2 \in \delta(q_0, s_2)$ such that
$q_1 \in \delta(q_2, s_3)$ with $s'=s_2s_3$ and $q_2$ violates the
controllable condition. Then, we have $s_1=s_2s_3$ and $\sigma_1
\notin \Gamma(q_2)$, where $\sigma_1 \in \Sigma_{uc}$ and
$s_2\sigma_1\in L(G)$. Thus, there is $q_2' \in \delta(q_0', s_2)$
such that $q_1' \in \delta(q_2', s_3)$ and $(q_2', q_2) \in
\phi_{1}$. Then $\sigma_1 \notin \Gamma'(q_2')$, which implies
that $R'$ does not satisfy the controllable condition. Hence, we
obtain a contradiction. Therefore, the assumption does not hold.
That is, $q_1 \in Q''$. Thus, $R'\prec R''$. Similarly, we can
prove that $R'\prec R''$ if $R'\prec_{\phi_{1}} R$ and $R'$ is
simulation-based strong observable. As a result, $R''$ is a
supremal element of $\mathcal{S}$.
\end{proof}

\begin{Remark}
The assumption requiring that $R\prec_{\phi} G$, can be satisfied
at the most cases because the descried specification should not be
out of the range of the behavior of the plant. This is similar to
the precondition $L(R)\subseteq L(G)$ in Ramadge-Wonham's
framework.
\end{Remark}

\section{EXAMPLE}
\begin{figure}[!htb]
\begin{center}
\includegraphics*[scale=.50]{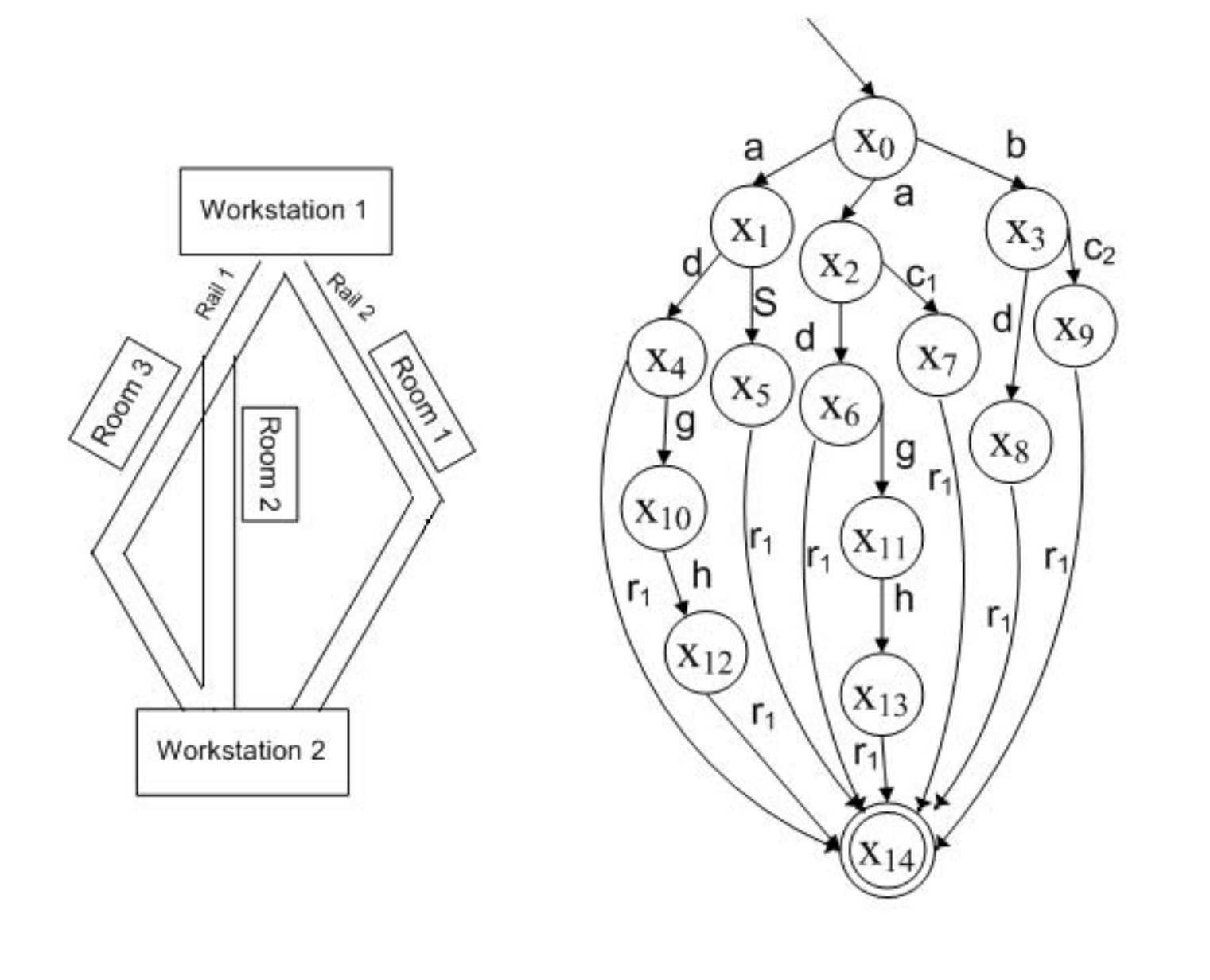}
\caption{Manufacturing System (Left) and Plant (Right)} \label{all3}
\end{center}
\end{figure}

Consider a manufacturing system that consists of two workstations,
three rooms and a robot as shown in Fig. \ref{all3} (Left).
Initially, the robot is in workstation 1. By choosing rail 1 (event
$a$), this robot nondeterministically goes to room 2 and room 3 and
by choosing rail 2 (event $b$), it can go to room 1. If the robot is
in room 2 and it hears the alarm (event $s$), it can go to the
workstation 2 (event $r_1$). Or it can take a video (event $d$) when
it is in room 2 and after that it has two choices : to go to
workstation 2 (event $r_1$) or to receive the message from the host
computer (event $g$). After the message has been received, the robot
can active an energy-saving mode (event $h$) and then go to
workstation 2 (event $r_1$). If the robot is in room 3, its behavior
is similar to what it does in room 2 except that it can pick up a
box from room 3 (event $c_1$) and then go to workstation 2 (event
$r_1$). If it is in room 1, it also has two choices: to pick up a
box from room 1 (event $c_2$) then go to workstation 2 (event $r_1$)
or to take a video (event $d$) and after then go to workstation 2
(event $r_1$). In this model, we assume that the event $s$
describing that the robot hears the alarm is uncontrollable, the
event $g$ describing that the robot receives a message from the host
computer is uncontrollable and unobservable and all the rest events
are controllable and observable.

\begin{figure}[!htb]
\begin{center}
\includegraphics*[scale=.50]{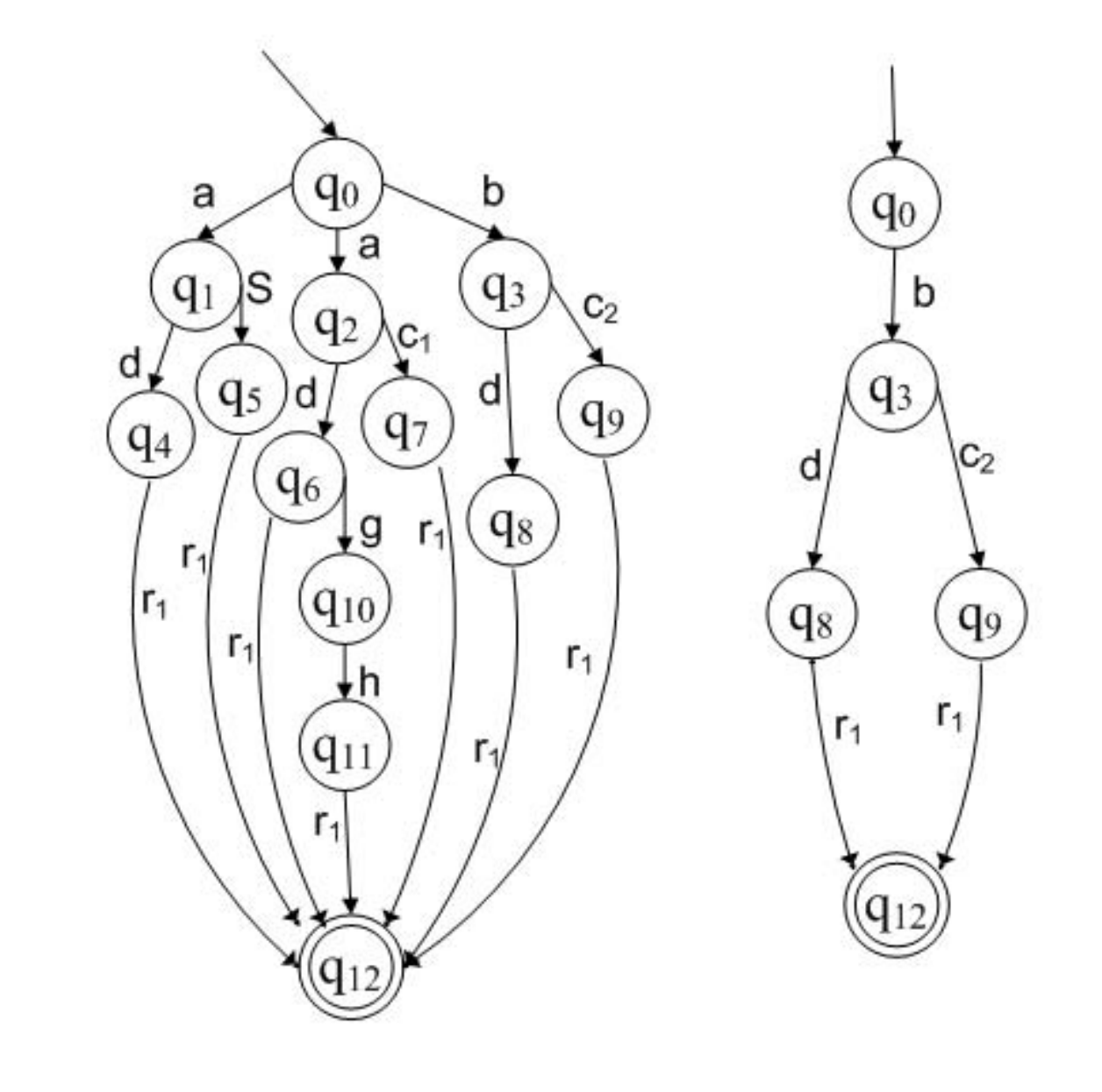}
\caption{Specification (Left) and Supremal Simulation-Based
Controllable and Strong Observable Subautomata (Right)} \label{all4}
\end{center}
\end{figure}


The automata model $G$ of the robot in manufacturing system is shown
in Fig. \ref{all3} (Right). The specification $R$ is in Fig.
\ref{all4} (Left) to restrict the behavior of $G$, which requires
that the robot can go to the workstation 2 after hearing the alarm
or go to workstation 2 after taking the video if it is in room 2. It
can be seen that $L(G)=L(R)$. Thus, if we use language equivalence
as a notion of behavioral equivalence, there is no need to control.
However, as mentioned above, $G$ can exhibit some undesired
behaviors, which motivates us to design a supervisor $S$ such that
the controlled system $S/G$ is bisimilar to $R$. In \cite{liu}, such
a supervisor $S$ exists if and only if $R$ is simulation-based
controllable and observable under partial observation. However, $R$
in this example is not simulation-based controllable and observable.
In this paper, we want to calculate the supremal simulation-based
controllable and strong observable subautomaton of $R$. By Algorithm
2, we obtain that $R$ is calculable for such kind of subautomaton.
Next, we have $q_2, q_4 \in Q_d(Q \times X)$, $q_1 \in Q_{d}'(Q
\times X)$ and $y_1=h_3(Q \times X)= \{(q_0, x_0), (q_3, x_3),
\{q_5, q_7, q_8, q_9, q_{11}\} \times \{x_4, x_5, x_6, x_7, x_8,
x_9, x_{12}, x_{13}\}, (q_6, x_4), (q_6, x_6), (q_{10}, x_{10}),
(q_{10}, x_{11}), (q_{12}, x_{14})\}$ in the first iteration.
Further, $y_2=h_3(y_1)=y_1$ and $(q_0, x_0) \in y_2$. Hence, the
supremal simulation-based controllable and strong observable
subautomata is achieved in Fig. \ref{all4} (Right).

\section{CONCLUSIONS}

By resorting to lattice theory, we proposed a computational approach
to solve the supremal simulation-based controllable and strong
observable subautomata, where both plant and specification are
modeled as nondeterministic automata. The obtained solution provides
a sufficient condition of the existence of the supremal
simulation-based controllable and strong observable subautomta and
an explicit algorithm to calculate such subautomta. Further, an
example is generated to illustrate the proposed techniques.





\end{document}